 \documentclass[final,authoryear,5p,times,twocolumn]{elsarticle}

\usepackage{graphicx}

\usepackage{amssymb}

\usepackage{rotating}
\usepackage{array}
\usepackage{color}
\usepackage{multirow}
\usepackage{float}
\usepackage{url}
\usepackage{ulem}
\usepackage{subfigure}
\usepackage{hyperref}
\hypersetup{
bookmarks=true,        
unicode=false,          
pdftoolbar=true,        
pdfmenubar=true,        
pdffitwindow=false,     
pdfstartview={FitH},    
pdftitle={N2 emission on titan},    
pdfauthor={Sonal Kumar Jain},   
pdfsubject={N2 triplet emission on Titan},   
pdfcreator={TeXlive},   
pdfproducer={TeXmakerX}, 
pdfnewwindow=true,      
colorlinks=false,        
linkcolor=blue,         
linkcolor=blue,         
citecolor=blue,        
filecolor=magenta,      
urlcolor=red            
}

\journal{J. Geophys. Research}

\begin{document}

\begin{frontmatter}



\title{\bf CO Cameron band and CO$_2^+$ UV doublet emissions in the dayglow of Venus: Role of CO
in the Cameron band production}
\author{Anil Bhardwaj} 
\cortext[cor1]{Corresponding author. Fax: +91 471 2706535 } 
\ead{anil\_Bhardwaj@vssc.gov.in; bhardwaj\_spl@yahoo.com}
\author{Sonal Kumar Jain \corref{cor1}}
\ead{sonaljain.spl@gmail.com}
\address{Space Physics Laboratory,
Vikram Sarabhai Space Centre,
Trivandrum~695022, India}

\begin{abstract}
Present study deals with the model calculations of CO Cameron band  
and CO$_2^+$ ultraviolet doublet emissions in the dayglow of Venus.
The overhead and limb intensities of CO Cameron band and CO$_2^+$ UV doublet
emissions are calculated for low, moderate, and high solar activity conditions.
Using updated cross sections, the impact of different e-CO cross section for Cameron 
band production is estimated.
The electron impact on CO is the major source mechanism of Cameron band,
followed by electron and photon impact dissociation of CO$_2$.
The overhead intensities of CO Cameron band and CO$_2^+$ UV
doublet emissions are about a factor of 2 higher in solar maximum
than those in solar minimum condition.
The effect of solar EUV flux models on the emission intensity  is $ \sim $30-40\% in solar 
minimum condition and $\sim$2-10\% in solar maximum condition.
At the altitude of emission peak ($ \sim $135 km), the model predicted limb intensity of
CO Cameron band  and CO$_2^+$ UV doublet emissions in moderate (F10.7=130) solar activity
condition is about 2400 and 300 kR, respectively, which is in agreement with the very recently
published SPICAV/Venus Express observation. The model limb intensity profiles of CO Cameron
band and CO$_2^+$ UV doublet are compared with SPICAV observation.
We also calculated intensities of N$_2$ Vegard-Kaplan UV bands and 
OI 2972 \AA\ emissions during moderate and high solar activity conditions.
\end{abstract} 

\end{frontmatter}

\section{Introduction}\label{sec:intro}
Several spacecraft, viz., Mariner 5 (3-channel photometer: 1050--2200 \AA, 1250--2200 \AA,
1350--2200 \AA); Venera 4 (1050--1340 \AA, 1225--1340 \AA); Mariner 10 (200--1700 \AA); 
Venera 9  and 10 (visible spectrometers 3000--8000 \AA{} and Lyman $ \alpha $
filter photometer); Venera 11 and 12 (300--1700 \AA);  Pioneer Venus Orbiter
(1100--1800 \AA\ and 1600--3300 \AA); Cassini (561--1182 \AA\ and 1155--1913 \AA) have
visited Venus so far. A review of past observations of Venus missions is given by \cite{Fox91}.
Currently, the Venus Express (VEx) is orbiting Venus which has an experiment
(SPICAV, 1100--3100 \AA, 7000--17000 \AA, 23000--42000 \AA) for aeronomical
studies of Venusian atmosphere. The major emission detected in the dayglow of Venus
include HeI 584 \AA{} and HeII 304 \AA\ lines, OI 989 \AA, OI 1304 \AA\ triplet, OI 1356 \AA,
OI 2972 \AA, OII 834 \AA,  CI 1561 and 1657 \AA, H Lyman-$\alpha$,
and CO fourth positive and Hopfield--Birge bands
\citep[e.g.,][]{Broadfoot74,Broadfoot77,Bertaux81,LeCompte89,Gerard11,Gerard11b,Hubert12}.

Theoretical calculations have shown CO Cameron band to be one of the brightest features 
(18--20 kR overhead intensity for low solar activity condition) in the UV  dayglow  of Venus 
\citep{Fox81,Fox91,Gronoff08}. The production sources of CO Cameron band ($a^3\Pi - X^1\Sigma^+$)
on Venus are expected to be similar to those on Mars, viz., photon and electron impact
dissociation of CO$_2$, dissociative recombination (DR) of CO$_2^+$, and electron impact on CO (e-CO).
Since $X^1\Sigma^+ \rightarrow a^3\Pi$ is a forbidden transition, resonance fluorescence
of CO is not an effective excitation mechanism. The CO$_2^+$ UV doublet emission originate 
due to transition from  B$^2\Sigma_u^+$ state of CO$_2^+$ to the ground state  ($B^2\Sigma^+-X^2\Pi $).
Theoretical calculations predicted an overhead dayglow intensity of around 7--10 kR for UV doublet,
with photoionization being the dominant production mechanism \citep{Fox81,Gronoff08}.

The main objective of the present study is to understand the role of 
various processes governing the production of CO Cameron band  and CO$_2^+$ UV doublet emissions
in the dayglow of Venus in the light of updated cross sections and reaction rates.
Recently, we have developed a model for the CO Cameron band  and CO$_2^+$ UV doublet
emissions in the dayglow of Mars \citep{Jain12}. In the present study this
model is applied to Venus to calculate the CO Cameron band and CO$_2^+$ UV doublet
dayglow emissions for low, moderate, and high solar activity conditions.

After submission of this paper, the first observation of CO Cameron band
and CO$_2^+$ UV doublet emissions in the dayglow of Venus using the SPICAV aboard Venus
Express were reported \citep{Chaufray12}. Keeping the structure of the paper unchanged, we added a section
in the revised version, where we compared model prediction with the SPICAV observation.
Details of the model are given in Section~\ref{sec:model}, followed by 
results and discussion in Section~\ref{sec:result} and \ref{sec:discussion}, respectively.
The summary and conclusions are presented in Section~\ref{sec:summary}.

\section{Development of Model}\label{sec:model}
Primary photoelectron production rate is calculated using 
\begin{equation}\label{eq:per}
Q(Z,E)=  \sum_l n_l(Z) \sum_{j,\lambda}\sigma_l^I(j,\lambda) I(Z,\lambda) \
\delta \left(\frac{hc}{\lambda}-E-W_{jl}\right)
\end{equation}
\begin{equation}
I(Z,\lambda)=I(\infty,\lambda)\ exp\left[-\sec(\chi)\sum_{l} \sigma_l^{A}
(\lambda)\int_Z^{\infty}n_l(Z^{'})dZ^{'}\right]
\end{equation}
where $\sigma_l^A$ and $\sigma_l^I (j,\lambda)$  are the total photoabsorption 
cross section and the photoionization cross section of the $j$th ion state of
the constituent $l$ at wavelength $\lambda$, respectively;
$I(\infty,\lambda)$ is the unattenuated solar flux at wavelength $\lambda$, 
$n_l$ is the neutral density of constituent $l$ at altitude Z; $\chi$ is the solar zenith angle (SZA); 
$\delta(hc/\lambda-E-W_{jl})$ is the delta function, 
in which $hc/\lambda$ is the incident photon energy, W$_{jl}$ is 
the ionization potential of the $j$th ion state of the $l$th constituent, and $E$ is 
the energy of ejected electron.

To calculate the photoelectron flux we have adopted the Analytical Yield 
Spectra (AYS) technique \citep[cf.][]{Bhardwaj90b,Singhal91,Bhardwaj99a,Bhardwaj03}.
The AYS is an analytical representation of 
numerical yield spectra obtained using the Monte Carlo model 
\citep[cf.][]{Singhal80,Singhal91,Bhardwaj99d,Bhardwaj09}.
Using AYS the photoelectron flux has been calculated as 

\begin{equation}\label{eq:a}
\phi(Z,E)=\int_{W_{kl}}^{100\, \mathrm{eV}}\frac{Q(Z,E) U(E,E_0)}{{\displaystyle\sum_{l}} n_l(Z)\sigma_{lT}(E)} \ dE_0
\end{equation}
where $\sigma_{lT}(E)$ is the total inelastic cross section for the
$l$th gas with density  $n_l$, and $U(E,E_0)$ is the two-dimensional AYS,
which embodies the non-spatial information of degradation process. It 
represents the equilibrium number of electrons per unit energy at an
energy $E$ resulting from the local energy degradation of an incident
electron of energy $E_0$. For the CO$_2$ gas, the AYS is taken from \cite{Bhardwaj09},
and for other gases, viz., O$_2$, N$_2$, O, and CO, we have used the AYS
given by \cite{Singhal80}.
The ion and electron temperatures for solar minimum and
maximum conditions are taken from \cite{Fox09}.

\begin{figure}
\centering
\includegraphics[width=20pc]{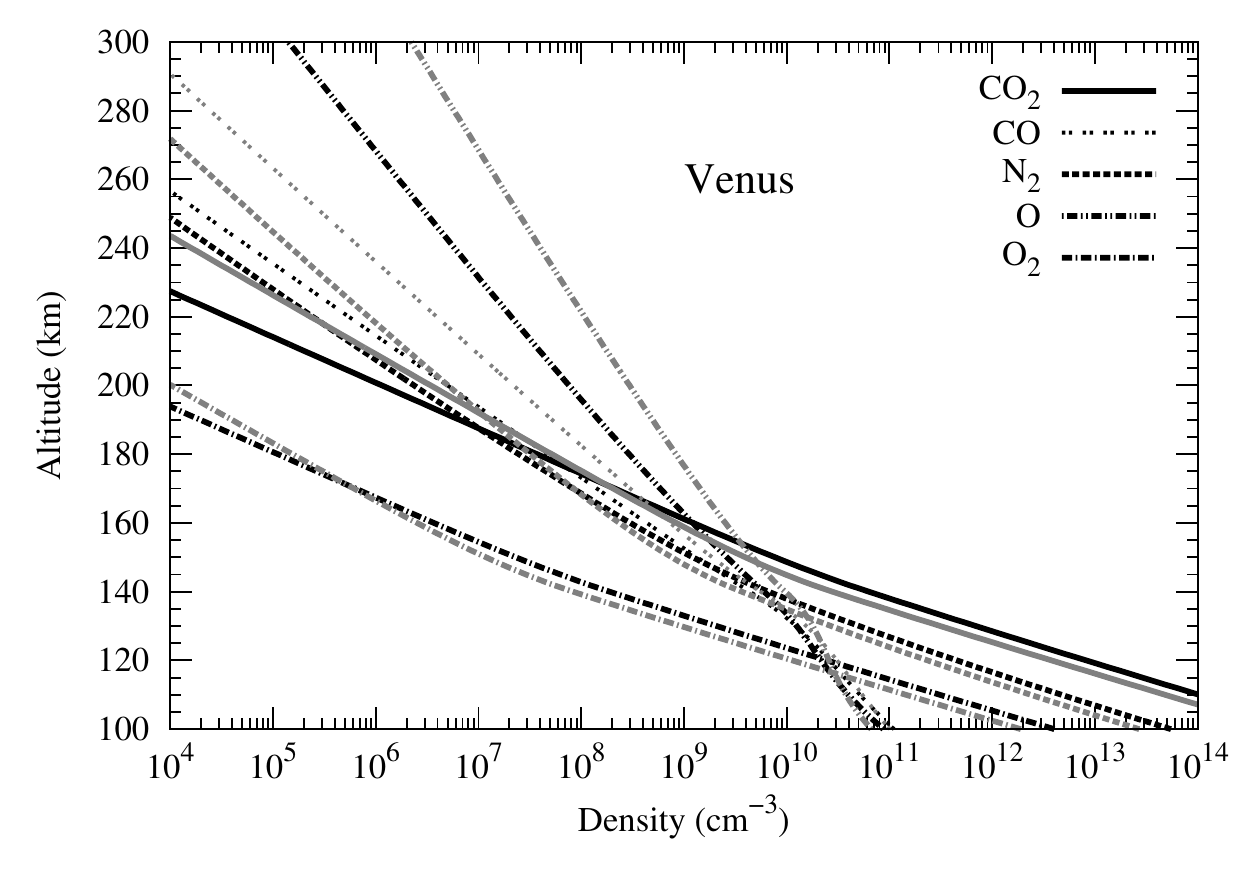}
\vspace*{-10pt}
\caption{Model atmosphere of  Venus for low (black curves)
and high solar (grey curves) activity conditions.}
\vspace*{-10pt}
\label{fig:atm-venus}
\end{figure}
Model atmosphere (considering CO$_2$, N$_2$, CO, and O) of Venus is taken from
the VTS3 model of \cite{Hedin83} for  solar minimum (F10.7 = 60), moderate (F10.7 = 130)
and  maximum (F10.7 = 200) conditions, for equatorial region
and local time of 1500 hrs, which corresponds to the solar zenith angle of around 45$^\circ$.
Based on the study of \cite{Fox91} the density of O$_2$ is taken as  $3 \times 10^{-3}$ times
that of CO$_2$  density. Figure~\ref{fig:atm-venus} shows the model 
atmosphere of Venus for low and high solar activity conditions. Below 160 km
(150 km in case of high solar activity) CO$_2$ is the major atmospheric species, but above
this altitude atomic oxygen becomes the dominant neutral in the atmosphere of Venus.

In the present study, solar EUV flux from  EUVAC model \citep{Richards94} is used for low, moderate,
and high solar activity conditions. To assess the impact of solar EUV flux model
on the calculated intensities, the solar EUV flux from SOLAR2000 (S2K) v2.36 model \citep{Tobiska04}
is also used. The solar EUV flux is taken at 1 AU and then scaled to the Sun-Venus
distance (0.72 AU).
There are substantial dif{}ferences in the solar EUV fluxes of EUVAC and S2K models;
moreover, these dif{}ferences are not similar in solar minimum and maximum conditions 
\citep[e.g., see][]{Jain12,Bhardwaj12b}. In both solar minimum and maximum conditions, the solar
flux estimated in bins is higher in S2K than in EUVAC over the entire
range of wavelengths, except for the bins below 250 \AA\ (150 \AA\ for solar minimum
condition), whereas solar fluxes at prominent lines are higher in EUVAC model for entire
wavelength range \citep[see Figure~1 of][]{Jain12}.
The higher solar fluxes above 250 \AA\ in S2K cause more photoionization.
Higher photon fluxes below 250 \AA\ (during solar maximum condition)
in EUVAC model produce more high-energy electrons causing
secondary ionizations that can compensate for the higher photoionization in S2K model.
A major dif{}ference between solar EUV flux of S2K and
EUVAC models is the solar flux at bin (1000--1050 \AA) containing
H~Ly-$ \beta $ (1026 \AA) line, which in both solar conditions is around an
order of magnitude higher in S2K compared to EUVAC solar flux model. The solar
fluxes at longer wavelength are very important for dissociative excitation processes.
Hence, contribution of photodissociation (PD) of CO$_2$  in CO(a$^3\Pi$)  production would be higher
when the S2K solar EUV flux model is used.

Due to its long lifetime, cross section for the production of CO(a$^3\Pi$) due to
electron impact dissociation of CO$_2$ (e-CO$_2$) is dif{}ficult to measure in the laboratory.
\cite{Ajello71a} reported relative magnitudes of the cross section for the (0, 1) 
transition of CO Cameron band  at 215.8 nm and reported a value of 
$1 \times 10^{-17}$ cm$ ^{2} $ at 80 eV. 
\cite{Erdman83} later criticised the cross section obtained by \cite{Ajello71a} and advocated
a value of $ 9 \times 10^{-17} $  cm$ ^{2} $ at 80 eV due to a re-evaluation to 9 ms
of the radiative lifetime \citep{Johnson72}. They subsequently multiplied this value by a 
factor of 2.7 to account for higher mean velocity of CO(a$^3\Pi$) fragments, which 
might have escaped detection \citep{Wells72}. Therefore, \cite{Erdman83} reported 
a value of $ 2.4 \times 10^{-16} $ cm$ ^{2} $ at 80 eV.
\cite{Avakyan98} have estimated the CO Cameron band cross section based on the
cross section of \cite{Ajello71a} with the correction of \cite{Erdman83}.
\cite{Bhardwaj09} have analytically fitted the cross section of CO(a$^3\Pi$) production
due to electron impact on CO$_2$ using the suggested value of \cite{Erdman83}.

\begin{figure}
\centering
\includegraphics[width=20pc]{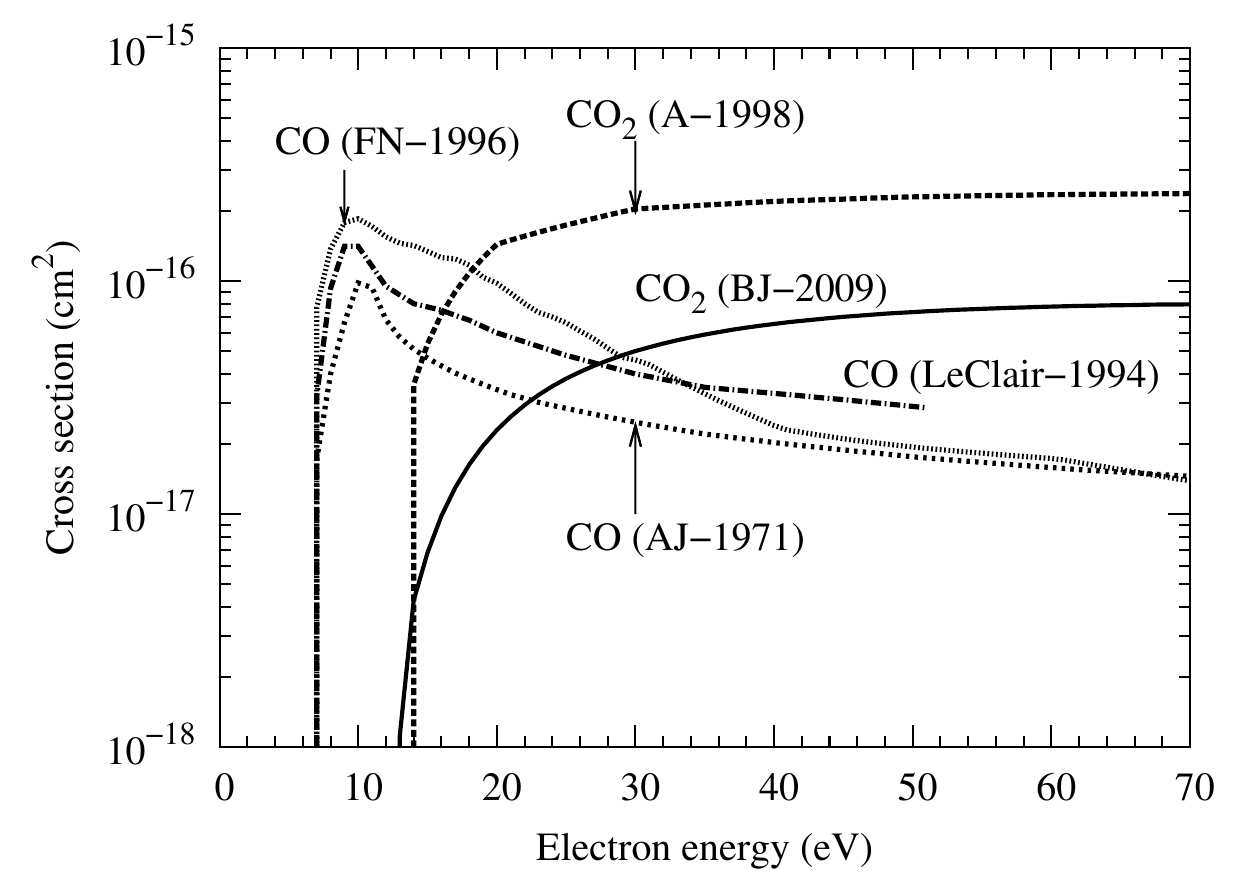}
\vspace*{-5pt}
\caption{Cross sections for the production of CO(a$^3\Pi$) due to electron impact on
CO and CO$_2$. A-1998, BJ-2009, FN-1996, LeClair-1994 and AJ-1971 refer to \cite{Avakyan98},
\cite{Bhardwaj09}, \cite{Furlong96}, \cite{LeClair94b}, and \cite{Ajello71b}, respectively. BJ-2009
cross section is plotted after dividing it by a factor of 3.}
\label{fig:cameron-xs}
\end{figure}

\cite{Conway81} constructed a synthetic spectrum of Martian dayglow between 1800 and 2600 \AA.
Based on the comparison of the model calculation with Mariner observation, \citeauthor{Conway81}
found that a cross section with a maximum value of $7 \times 10^{-17}$ cm$^{2}$ was
consistent with the data. The value suggested by \cite{Conway81} is around  a factor
of 3 smaller than that of \cite{Erdman83}.
Recent comparison between calculations and observations of dayglow emission on
Mars suggests a lower value of e-CO$_2$ cross sections for the CO Cameron band
production  \citep{Simon09,Jain12,Gronoff12}. \cite{Jain12} and \cite{Gronoff12}
have shown that Cameron band cross sections  of \cite{Erdman83} should be reduced
by a factor of 2 to 3, to bring the calculated CO Cameron band intensities in
agreement with the Mars Express observation.
The reduction in the CO(a$^3\Pi$) cross section is also supported by recent measurements
of radiative lifetime of CO(a$^3\Pi$).
Based on theoretical and experimental work, \cite{Gilijamse07} have re-analysed the radiative
lifetime of CO(a$^3\Pi$), and reported a value of $ \sim $3.16 ms, which is around 3 times less
than the value of \cite{Johnson72}.
In the present study the cross section for CO(a$^3\Pi$) production due to electron impact on CO$_2$ is taken from 
\cite{Bhardwaj09} after dividing it by a factor of 3, which is shown in
Figure~\ref{fig:cameron-xs} along with the recommended cross section of \cite{Avakyan98}.

Electron impact on CO (e-CO) is also a source of CO Cameron band. On Mars, due
to less abundance of CO, it does not contribute significantly to the total Cameron band
emission \citep{Fox79,Jain12}. However, on Venus, CO contribution  cannot be neglected
due to its relatively larger abundance above 150 km (cf. Figure~\ref{fig:atm-venus}).
In comets, where the CO abundance is larger or equal to that of the CO$_2$, the major contribution to CO
Cameron band emission is from electron impact on CO \citep{Bhardwaj11a,Raghuram12}.
In the previous calculations of CO Cameron band emission in the dayglow of
Venus \citep{Gronoff08,Fox81,Fox91}, the e-CO cross section for CO(a$^3\Pi$) production
was taken from the work of \cite{Ajello71b}.
\cite{Ajello71b} used the (1,4) Cameron band at 2389 \AA\ to normalize the entire
band system cross section in electron impact excitation of CO. However, according
to  \cite{Erdman83}, the (1,4) Cameron band was contaminated by (6,16) CO fourth
positive band. \cite{Erdman83} repeated and re-analysed the \citeauthor{Ajello71b}'s
experiment with higher sensitivity and concluded that total cross section
value  ($ 1.1 \times 10^{-16} $ cm$^{2}$ at 11 eV) measured by \cite{Ajello71b} should, therefore,
be reduced by a factor of 8 to an apparent value of $ 1.4 \times 10^{-17} $ cm$^{2}$ at 11 eV.
In addition to the contamination problem, \citeauthor{Ajello71b}'s total Cameron band
emission cross section was based on the assumption of radiation lifetime of 1 ms for
a$^3\Pi$ state. \cite{Erdman83} used the radiative life of 9 ms \citep{Johnson72}
and multiplied the cross section (already corrected for contamination) by a factor of 9
and gave a cross section value of $ 1.5 \times 10^{-16} $ cm$^{2}$ at 11 eV, with an
uncertainty close to 75\%.

After accounting for corrections, the cross section value suggested by \cite{Erdman83}
is very close to the cross section of \cite{Ajello71b}. However, based on
CO(a$^3\Pi$) radiative lifetime of $\sim$3 ms reported by  \cite{Gilijamse07},
the Cameron band cross section in e-CO process should be reduced by 
a factor of 3.

\cite{LeClair94b} have measured the e-CO cross 
section for CO(a$^3\Pi$) production using solid xenon detector and time of flight (TOF) technique.
\cite{LeClair94b} have given the integral cross section (ICS) of CO(a$^3\Pi$)---that
include cascading contributions from higher triplet states---by normalizing
their excitation function to the maximum absolute cross section ($ 1.5 \times 10^{-16} $
cm$^{2}$ at 11 eV) obtained by \cite{Erdman83}. This normalization may cause an overestimation
of CO(a$^3\Pi$) section measured by \cite{LeClair94b}, since \cite{Erdman83} have used
the radiative lifetime of 9 ms, which is a factor of 3 higher than recently measured
lifetime of 3 ms \citep{Gilijamse07}.
The shape of normalized CO(a$^3\Pi$) cross section measured
by \cite{LeClair94b} is identical to the one recorded by \cite{Ajello71b}. However, maximum
cross section is at 9.4 eV in \cite{LeClair94b} measurement compared to 11 eV in \cite{Ajello71b}
experiment. \cite{LeClair94b} attributed this difference in peak position to the electron
beam characteristic in the two experiments.

\cite{Furlong96}  reported the absolute integral cross section for CO(a$^3\Pi$)
production in the e-CO collision by normalizing their measurements to maximum
cross section value  ($1.698 \times 10^{-16}$ cm$^2$ at 8.5 eV)  calculated by  \cite{Morgan93}.
Below 10 eV, their cross section is in good agreement with that of \cite{LeClair94b}.
Above 10 eV, \cite{Furlong96} reported an increase in cross section due to the
contribution from cascading into a$ ^3\Pi $ state.
The cross sections obtained by \cite{Furlong96} are about a factor
of 2 higher between 10 and 35 eV compared to that of \cite{LeClair94b}.

The above mentioned discussion clearly points out the difference in the cross
section of CO(a$^3\Pi$) in electron impact excitation of CO.
In the present study, cross section of \cite{Furlong96} is used for
CO(a$^3\Pi$) production in e-CO collision. The cross section of \cite{LeClair94b} is
also used to assess the effect of cross section in Cameron band intensity.
The reason for using these two cross sections over the  one measured by \cite{Ajello71b}
is due to the fact that \citeauthor{Ajello71b}'s measured cross section have been shown
to be flawed by \cite{Erdman83}. Figure~\ref{fig:cameron-xs} depicts the CO(a$^3\Pi$) cross
sections in e-CO process used in the present study along with cross section obtained by
\cite{Ajello71b}. The cross section of CO(a$^3\Pi$) production in e-CO process attains
maximum value at $ \sim $10 eV, where photoelectron flux also has high values \citep{Bhardwaj12b}. This
makes e-CO collisions more important for the Cameron band production, if CO density
is sufficient, as in the case of Venus. At electron energies $ > $25 eV, CO(a$^3\Pi$) 
cross section in e-CO$_2$ process becomes dominant (cf. Figure~\ref{fig:cameron-xs}).

\begin{figure}
\centering
\includegraphics[width=20pc]{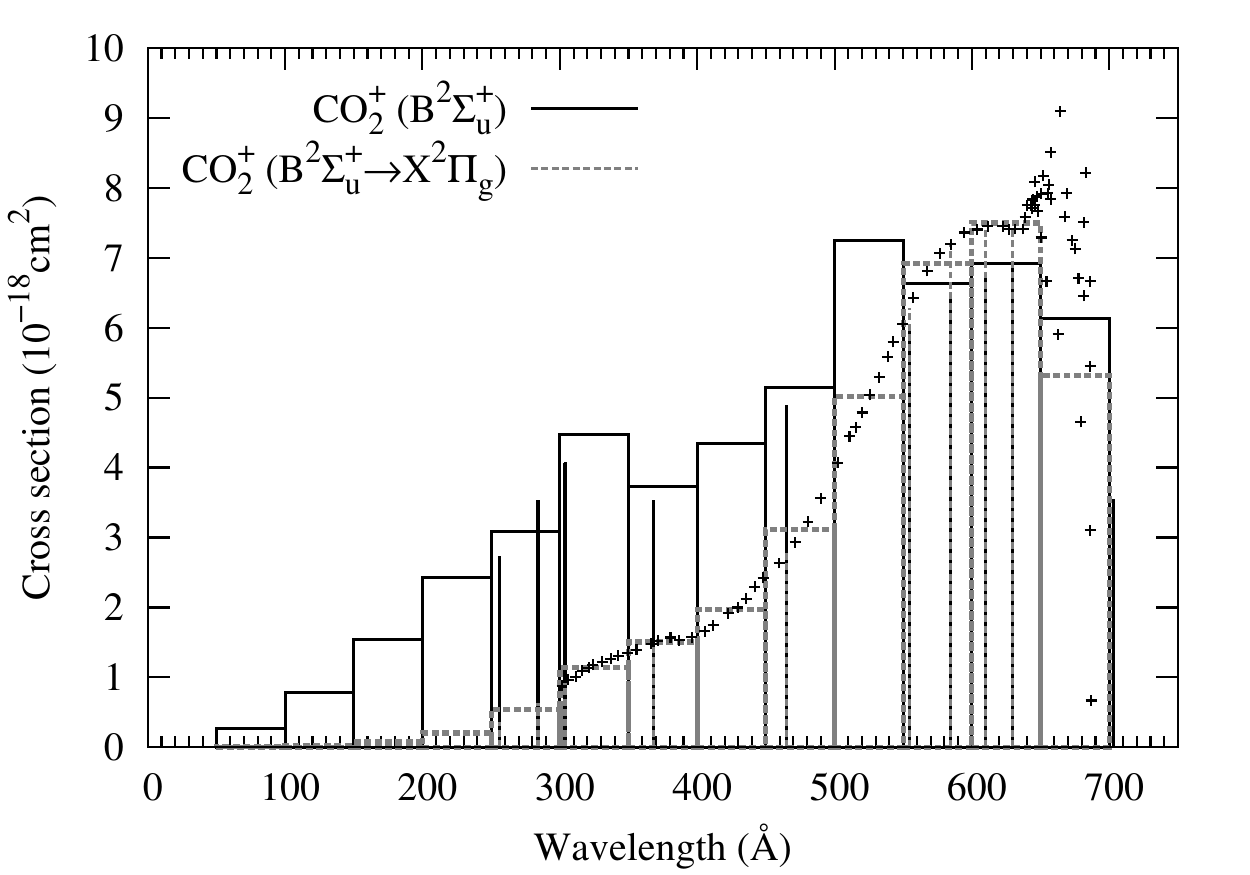}
\caption{Photon impact excitation cross section 
of CO$ _2^+$(B$^2\Sigma_u^+$) taken from \cite{Schunk00} with branching ratio from \cite{Avakyan98} (black lines).
Symbols show the emission cross section of CO$_2^+$(B$^2\Sigma_u^+$ $ \rightarrow $ X$^2\Pi_g $) as given in \cite{Ukai92}.
Grey lines show the emission cross section of CO$_2^+$(B$^2\Sigma_u^+$ $ \rightarrow $ X$^2\Pi_g $)
averaged at 37 wavelengths bins; at wavelengths below 300 \AA, cross section is extrapolated.}
\label{fig:co2buv-xs}
\end{figure}
The details of photoabsorption, photoionization, and electron impact cross
sections used in the present study are given in our previous work \citep{Jain11,Jain12,Bhardwaj12b}.
The details of cross sections and processes considered in the model to calculate  CO
Cameron band and CO$_2^+$ UV doublet emissions are summarised in Table~\ref{tab:xs-summary}.
While calculating the emission from B$ ^2\Sigma_u^+ $ state of CO$_2^+$, 
we have taken branching ratio of 0.5 (for photoionization only)
from the CO$_2^+$ (B)  to (A) based on the study of \cite{Fox79}. \cite{Ukai92} have
given the direct emission cross section of CO$_2^+$(B$^2\Sigma_u^+$ $ \rightarrow $ X$^2\Pi_g $)
transition. This cross section is also used in the present study to assess the
impact of using excitation and emission cross section of B$ ^2\Sigma_u^+ $ state of CO$_2^+$.
Figure~\ref{fig:co2buv-xs} shows the CO$_2^+$(B$^2\Sigma_u^+$) excitation
and CO$_2^+$(B$^2\Sigma_u^+$ $ \rightarrow $ X$^2\Pi_g $) emissions cross sections due to photon
impact on CO$ _2$. Emission cross section of \cite{Ukai92} has been averaged at
37 wavelength bins.
The contribution of fluorescence scattering of CO$_2^+$ to the
UV doublet emission is calculated by  taking the fluorescence efficiency  ({\it g})
value of  $ 5.2 \times 10^{-3} $ s$^{-1}$ for Venus \citep{Dalgarno71}.

\begin{table*}
\caption{CO(a$^3\Pi$)  and CO$_2^+(\mathrm{B}^2\Sigma_u^+)$ production processes and references for cross sections and reaction rates.}
\label{tab:xs-summary}
\begin{tabular*}{\textwidth}{@{\extracolsep{\fill}}ll} 
\hline  \noalign{\smallskip}
Process & References \\
\hline  \noalign{\smallskip}
CO$_2$ + e$_{ph}$ $\rightarrow$ CO(a$^3\Pi$) + O & \cite{Bhardwaj09} \\
CO$_2$ + h$\nu$ $\rightarrow$ CO(a$^3\Pi$) + O & \cite{Lawrence72a} \\
CO$_2^+$ + e$_{th}$ $\rightarrow$  CO(a$^3\Pi$) + O & \cite{Seiersen03,Skrzypkowski98} \\
CO + e$_{ph}$ $\rightarrow$ CO(a$^3\Pi$)  & \cite{Furlong96}\footnotemark[1] \\
CO$_2$ + e$_{ph}$ $\rightarrow$ CO$_2^+(\mathrm{B}^2\Sigma_u^+)$ & \cite{Bhardwaj09} \\
CO$_2$ + h$\nu$ $\rightarrow$ CO$_2^+(\mathrm{B}^2\Sigma_u^+)$ & \cite{Schunk00}\footnotemark[2] \\
CO$_2^+$ + h$\nu$ $\rightarrow$ CO$_2^+(\mathrm{B}^2\Sigma_u^+)$ & \cite{Dalgarno71} \\

\hline  \noalign{\smallskip}
\end{tabular*}
\footnotemark[1]{Cross section measured by \cite{LeClair94b} has also been used.}
\footnotemark[2]{Branching ratios for ionization in different states are from Avakyan et al. (1998).}
\end{table*}

\section{Results}\label{sec:result}
\subsection{Solar minimum condition}\label{subsec:smin-venus}
\begin{figure}
\centering
\includegraphics[width=20pc]{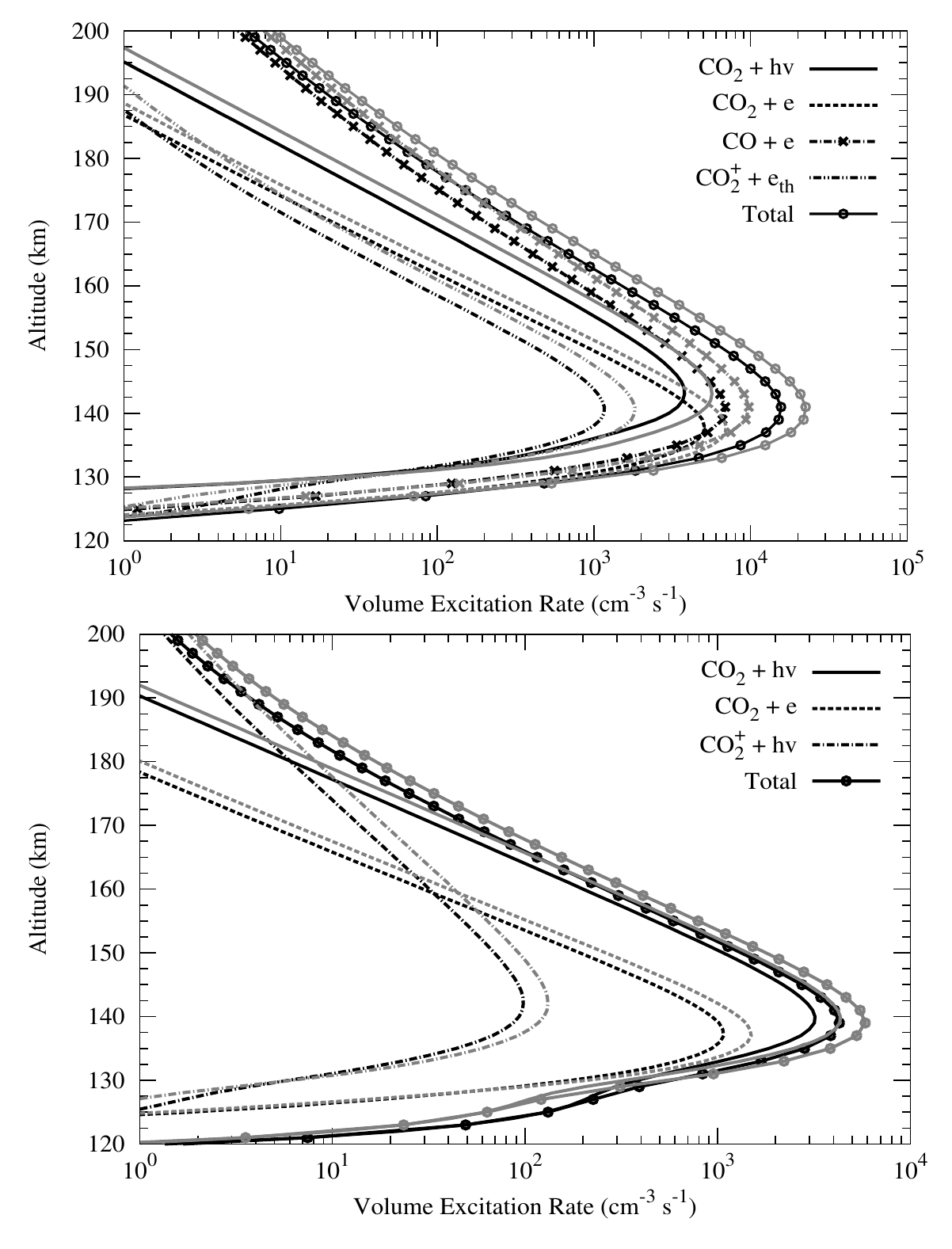}
\caption{Calculated production rates of the CO(a$^3\Pi$) (upper panel)  
and CO$_2^+(\mathrm{B}^2\Sigma_u^+)$ (bottom panel)
on Venus for low solar activity condition at SZA = 45$^\circ$. Black curves show production rates calculated 
using EUVAC model while grey curves show them for S2K solar flux model.}
\label{fig:ver-low-venus}
\end{figure}

Figure~\ref{fig:ver-low-venus} shows the calculated volume excitation rates of
CO(a$^3\Pi$) and CO$_2^+(\mathrm{B}^2\Sigma_u^+)$ for low solar activity condition.
The altitude of peak production is $\sim$140 km. 
The major production source of Cameron band at the peak is e-CO process, whose
contribution is about 44\%; unlike on Mars, where
electron impact on CO$_2$ is the major Cameron band production mechanism \citep{Jain12}.
Table~\ref{tab:oi-cmp-venus} shows the height-integrated overhead intensity
of CO Cameron band with contributions of different sources. The e-CO collisions  are the major
source of Cameron band production with contribution of around 45\%,
followed by e-CO$_2$, PD  of CO$_2$, and DR of CO$_2^+$, whose contributions are around 25,
23, and 7\%, respectively.
\begin{table*}
\caption{Overhead intensities (in kR) of CO Cameron  band and CO$_2^+$ UV doublet emissions on Venus for
low, moderate, and high solar activity conditions  at  SZA = 45$^\circ$.}\label{tab:oi-cmp-venus}
\begin{tabular*}{\textwidth}{@{\extracolsep{\fill}}llllllll}
\hline  \noalign{\smallskip}
Process 	& \multicolumn{7}{c}{Intensity (kR)} \\
\cline{2-8}   \noalign{\smallskip}
&	\multicolumn{3}{c}{CO Cameron Band} &	& \multicolumn{3}{c}{CO$_2^+$ UV doublet} \\  \noalign{\smallskip}
\cline{2-4} \cline{5-8}   \noalign{\smallskip}
& Low SA\footnotemark[1]	& Mod. SA\footnotemark[2]	& High SA\footnotemark[3] & & Low SA & Mod. SA	& High SA\\  \noalign{\smallskip}
\hline  \noalign{\smallskip}
&	\multicolumn{7}{c}{EUVAC solar flux model} \\  \noalign{\smallskip}
\cline{2-8}  \noalign{\smallskip}
CO$_2$  + h$\nu$	& 5.7  (6.2)\footnotemark[4]& 7.2 &7.5 && 4.8 &8.5	& 9.5	\\
e$^-_{ph}$ + CO$_2$		& 6.6  (7.8) & 12.3	& 13.7 				&& 1.4 &2.7	& 3		\\
e$^-_{ph}$ + CO			& 11.4 [7.8]\footnotemark[5] (2.9)& 27.3	& 36.3 [25.6]			&& -   &-	& -		\\
e$^-_{th}$ + CO$_2^+$ & 1.7 (2)	& 2.9	& 2.9		&& -   &-	& -		\\
FS						&			&	-	& -					&& 0.2 &0.3	& 0.3	\\
Total			& 25.3 [21.8] (18)	& 49.8	& 60.4	[49.8]		&& 6.4 \{4\}\footnotemark[6] &11.5 \{7.2\}& 12.8 \{8\}	\\[5pt]
&	\multicolumn{7}{c}{SOLAR2000 solar flux model} \\ \noalign{\smallskip}
\cline{2-8}  \noalign{\smallskip}
CO$_2$  + h$\nu$	& 8.6 		&10.7	&	11.6		&& 6.3 &8	& 8.7	\\
e$^-_{ph}$ + CO$_2$		& 9.2		&11		&	11.7		&& 1.9 &2.4	&2.5  	\\
e$^-_{ph}$ + CO			& 16.2 		&26.3	&	33.5		&& -   &-	& -		\\
e$^-_{th}$ + CO$_2^+$		& 2.6 		&2.8	&	2.6			&& -   &-	& -		\\
FS				&-			&-		&-				&& 0.4 &0.3	& 0.3	\\
Total			& 36.3		&51		&	59.4		&& 8.6 \{5.5\} &10.7 \{6.7\}& 11.4 \{7.2\}	\\
\hline
\end{tabular*}
{\small e$^-_{ph}$ = Photoelectron;  e$^-_{th}$ = Thermal electron; FS = Fluorescent  scattering of CO$_2^+$}\\
\footnotemark[1]{Low solar activity (F10.7=60).} 
\footnotemark[2]{Moderate solar activity (F10.7=130).}
\footnotemark[3]{High solar activity (F10.7=200).}
\footnotemark[4]{Calculated values in parenthesis are for model atmosphere of \cite{Fox81} and e-CO Cameron 
band production cross section from \cite{Ajello71b}.}
\footnotemark[5]{Calculated values in brackets are for the CO(a$^3\Pi$) cross section of \cite{LeClair94b}} 
\footnotemark[6]{Calculated by taking the 50\% cross-over from $ B $ to $ A $ before radiating.}
\end{table*}

\begin{figure*}
\centering
\includegraphics[width=\textwidth]{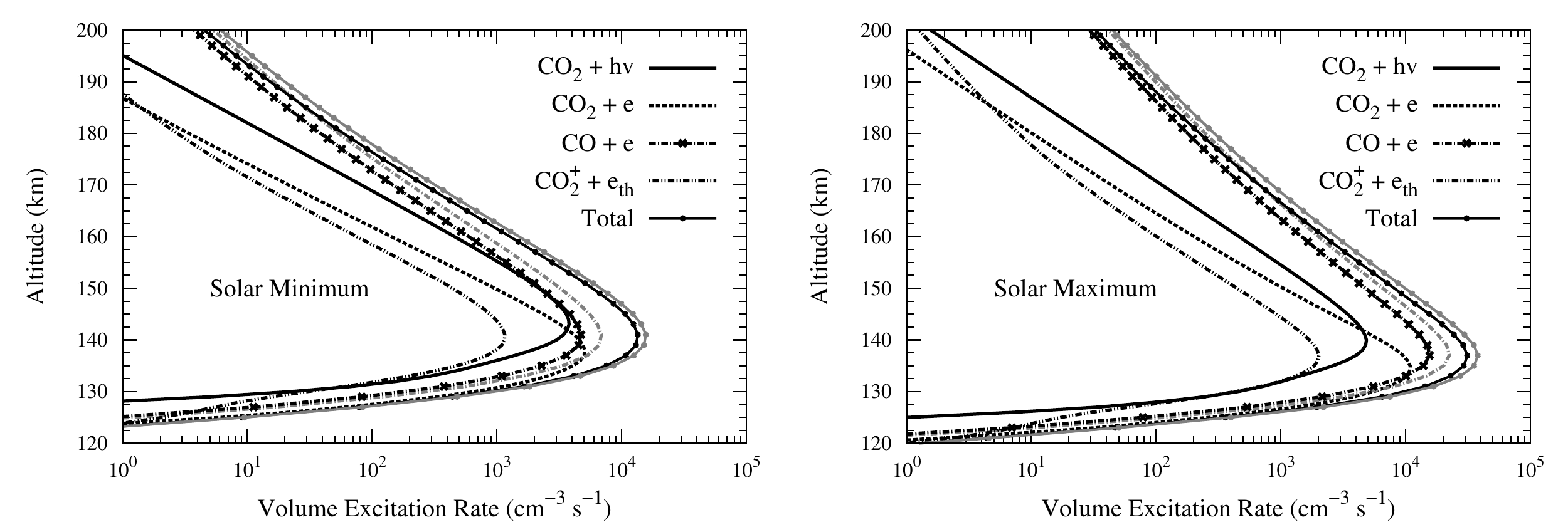}
\caption{Calculated production rates of the CO(a$^3\Pi$) on Venus for low (left panel)
and high (right panel) solar activity conditions at SZA = 45$^\circ$. Black curves show production rates calculated 
using EUVAC model and the CO(a$^3\Pi$) cross section in e-CO process from \cite{LeClair94b}, while grey
curves show the production rate of CO(a$^3\Pi$) in e-CO process and total production rate when CO(a$^3\Pi$)
cross section are taken from \cite{Furlong96}.}
\label{fig:ver-cameron-xs}
\end{figure*}
Bottom panel of Figure~\ref{fig:ver-low-venus} shows the volume production 
rate of CO$_2^+(\mathrm{B}^2\Sigma_u^+)$, and the height-integrated overhead
intensity of CO$_2^+$ UV doublet emission for different sources is presented
in Table~\ref{tab:oi-cmp-venus}. Photoionization of CO$_2$ 
is the dominant source (75\%) of CO$_2^+(\mathrm{B}^2\Sigma_u^+)$ production,
followed by the electron impact ionization of CO$_2$ (21\%). Contribution of
fluorescent scattering of CO$_2^+$ is very small ($ \sim $3\%).

Figure~\ref{fig:ver-cameron-xs} shows the volume production rate of CO(a$^3\Pi$) for
low  and high solar activity conditions calculated by using the CO(a$^3\Pi$) cross section
measured by \cite{LeClair94b}. The height-integrated intensity of Cameron band
is given in Table~\ref{tab:oi-cmp-venus}. The e-CO process is still the dominant
source of Cameron band production, though its contribution in Cameron band production
is reduced compared to the case when CO(a$^3\Pi$) cross section is taken from \cite{Furlong96},
which is consistent with our previous considerations.

The volume excitation rates are integrated along the line of sight to calculate
the limb intensities of CO$_2^+$ UV doublet and CO Cameron band emissions
in the dayglow of Venus. Limb intensity at each tangent point is calculated as 

\begin{equation}\label{eq:los}
	I = 2 \int_{0}^{\infty} \mathrm{V}(r)dr,
\end{equation}
where $r$ is abscissa along the horizontal line of sight, and V(r) is the volume
emission rate (in cm$^{-3}$ s$^{-1}$) at a particular emission point $ r $.
The factor of 2 multiplication comes due to symmetry along the line of sight 
with respect to the tangent point. While calculating limb intensity we assumed
that the emission rate is constant along local longitude/latitude.
Figure~\ref{fig:int-low-venus} shows the
limb intensities of CO$_2^+$ UV doublet and CO Cameron band emission
on Venus. The calculated limb intensity of Cameron band peaks at 137 km with a
value of 1200 kR, while the maximum limb intensity of CO$_2^+$ UV doublet emission is 183 kR at
an altitude of 136 km.

\subsection{Solar maximum condition}\label{subsec:smax-venus}
Figure~\ref{fig:ver-max-venus} shows the calculated volume excitation rates
of CO Cameron band (upper panel)  and CO$_2^+$ UV doublet (lower panel)  emissions
for solar maximum condition. The production rate of Cameron band attains a 
maximum value of $ 3.8 \times 10^4 $ cm$ ^{-3} $ s$ ^{-1} $ at an altitude
of 137 km. The height-integrated overhead intensity is presented in
Table~\ref{tab:oi-cmp-venus}. Electron impact on CO is by far the dominant
production source of Cameron band contributing about 60\%, followed by
electron impact on CO$_2$ (23\%), PD of CO$_2$ (12\%), and DR of CO$_2^+$ (4\%).
The CO(a$^3\Pi$)  production rate calculated using e-CO cross section from \cite{LeClair94b}
is shown in Figure~\ref{fig:ver-cameron-xs} and corresponding height-integrating
intensities in Table~\ref{tab:oi-cmp-venus}.

For the CO$_2^+$ UV doublet emission, maximum production rate occurs at an altitude
of 135 km with a value of $\sim8.7 \times 10^3$ cm$^{-3}$ s$ ^{-1}$ 
(cf. Figure~\ref{fig:ver-max-venus}). The  overhead intensity of CO$_2^+$ UV doublet
is presented in  Table~\ref{tab:oi-cmp-venus}. The PD of CO$_2$  is
the dominant (74\%) production source of UV doublet emission followed by
electron impact on CO$_2$ (23\%) and fluorescent scattering by CO$_2^+$ (3\%).
Figure~\ref{fig:int-max-venus} shows the calculated line of sight intensities
of CO Cameron band and CO$_2^+$ UV doublet emissions.
The intensity of Cameron band peaks $\sim$135 km with a value of 2700 kR, while the intensity
of UV doublet emission attains a maximum value of around 380 kR at an altitude
of 132 km.

\subsection{Solar moderate condition}\label{subsec:smod-venus}
The model calculation is also carried out for the moderate solar activity 
condition by taking the solar EUV flux on 1 July 2012 (F10.7 = 130).
The height-integrated overhead intensities of CO Cameron band and CO$_2^+$ UV
doublet emissions are presented in Table~\ref{tab:oi-cmp-venus}. Our calculation
shows that for solar moderate condition also, the e-CO process is the dominant
mechanism of CO Cameron band production contributing about 55\%, followed by
electron impact on CO$_2$ (25\%), PD of CO$_2$ (14\%), and DR of CO$_2^+$ (6\%).
The PD of CO$_2$  is the dominant (74\%) production source of UV doublet emission
followed by electron impact on CO$_2$ (23\%) and fluorescent scattering by CO$_2^+$ (3\%).
Figure~\ref{fig:int-mod-venus} shows the calculated limb intensities of CO
Cameron band and CO$_2^+$ UV doublet emissions in the dayglow of Venus for
moderate solar activity condition. Both emissions maximise at $\sim$135 km
with intensity of $\sim$2200 kR for CO Cameron band and 330 kR for
CO$_2^+$ UV doublet emission.

\section{Discussion}\label{sec:discussion}
The present model calculation shows that the electron impact on CO is the dominant
source of CO Cameron band production in the atmosphere of Venus for
low, moderate, and high solar activity conditions using the CO(a$^3\Pi$) cross sections of
\cite{Bhardwaj09} and \cite{Furlong96} in electron impact on CO$_2$ and CO, respectively.
For solar minimum condition
\cite{Fox81} and \cite{Gronoff08} reported that  e-CO$_2$ process is the major
production source of Cameron band. \cite{Gronoff08} have calculated CO Cameron
band intensity of 17.3 kR; with 7 kR from electron impact on CO$_2$, 5.3 kR from
PD of CO$_2$, 4 kR from electron impact on CO, and 1 kR from DR of
CO$_2^+$. \cite{Gronoff08} have used the cross section of \cite{Ajello71b} for electron
impact on CO, while in the present study the cross section of \cite{Furlong96} has been 
used. Using the cross section of \cite{Ajello71b}, our model calculated 
overhead Cameron band  intensity is 18.6 kR, with contributions from  e-CO$_2$, PD
of CO$_2$, e-CO, and DR of CO$_2^+$ processes being  6.7, 5.6,
4.6, and 1.7 kR, respectively. The model calculated total CO Cameron band intensity is in 
good agreement with that of \cite{Gronoff08}.
\cite{Fox81} reported the Cameron band intensity of about 20 kR, with contribution of $ \sim $25\%
from DR of CO$_2^+$ and 6\% from  e-CO process. The present calculation, as well as that of  
\cite{Gronoff08}, show that the contribution of DR of CO$_2^+$ (which depends on electron density
and temperature) is smallest among the   processes considered in the model 
(see Table~\ref{tab:oi-cmp-venus}).
\cite{Fox91}  suggested that the source of DR was overestimated in the pre-Pioneer
Venus model of \cite{Fox81} because of low density of  atomic oxygen, which led to larger
densities of CO$_2^+$ ion.
The mixing ratio of CO was lower in the model atmosphere used by  \cite{Fox81}, whereas
in the present calculation, as well as in the model of \cite{Gronoff08}, the VTS3 model 
atmosphere is used, which has larger CO mixing ratio.
To evaluate the effect of low CO mixing ratio, the model calculation
is also carried out by taking model atmosphere of \cite{Fox81}; the
results are shown in Table~\ref{tab:oi-cmp-venus}. The Cameron
band intensity is 18 kR when the model atmosphere of \cite{Fox81} and
e-CO cross section of \cite{Ajello71b} are used, which is in
agreement with the model result of \cite{Fox81}. However,
in the present calculation the contribution of DR is about 11\%, which
is lower than that reported by \cite{Fox81}; this might be due
to the difference in DR rate coefficient for CO(a$^3\Pi$) production
in the two calculations.
\begin{figure}[h]
\centering
\includegraphics[width=20pc]{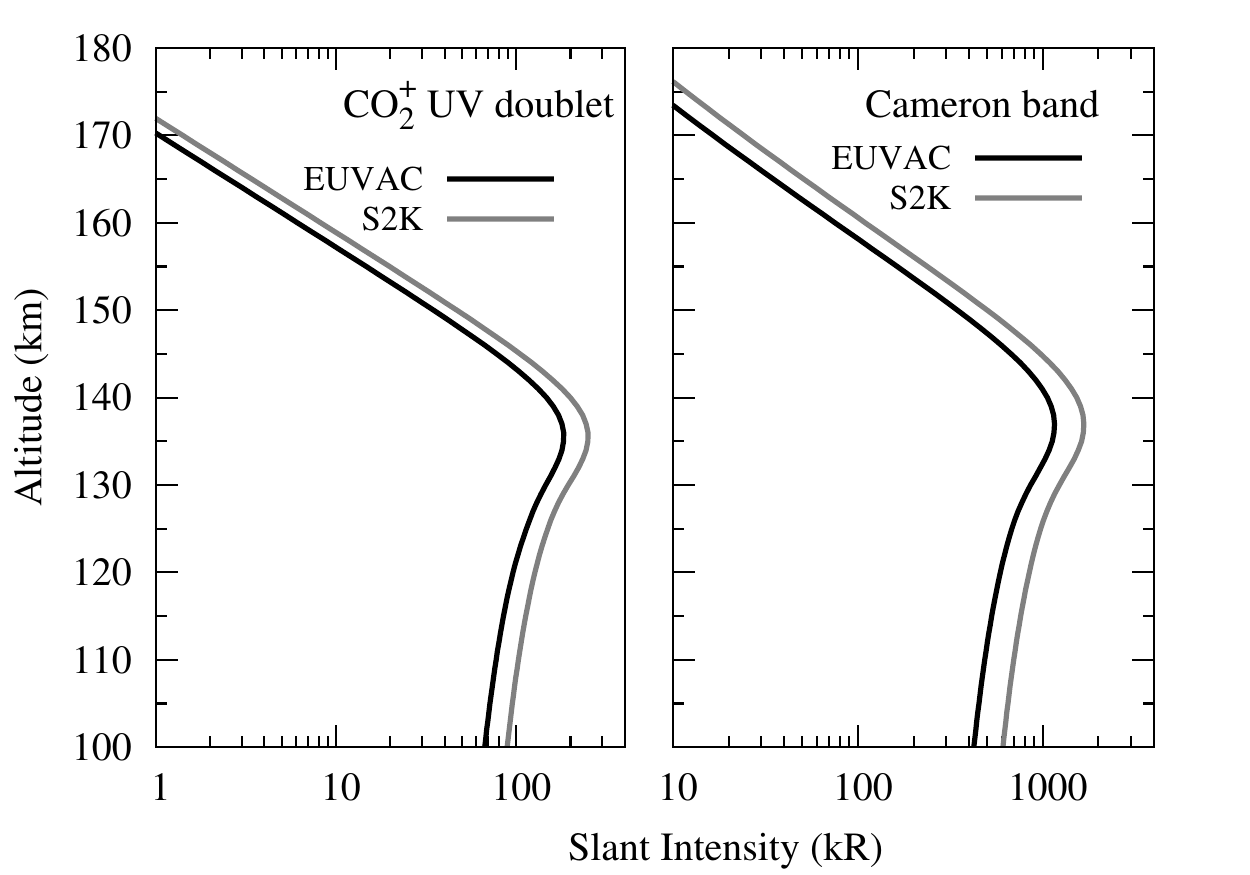}
\caption{Calculated limb profiles of CO$_2^+$ UV doublet (left panel) and
CO Cameron band emissions (right panel) for EUVAC and S2K solar EUV flux models for low solar activity condition
at SZA = 45$^\circ$.}
\label{fig:int-low-venus}
\end{figure}

\begin{figure}
\centering
\includegraphics[width=20pc]{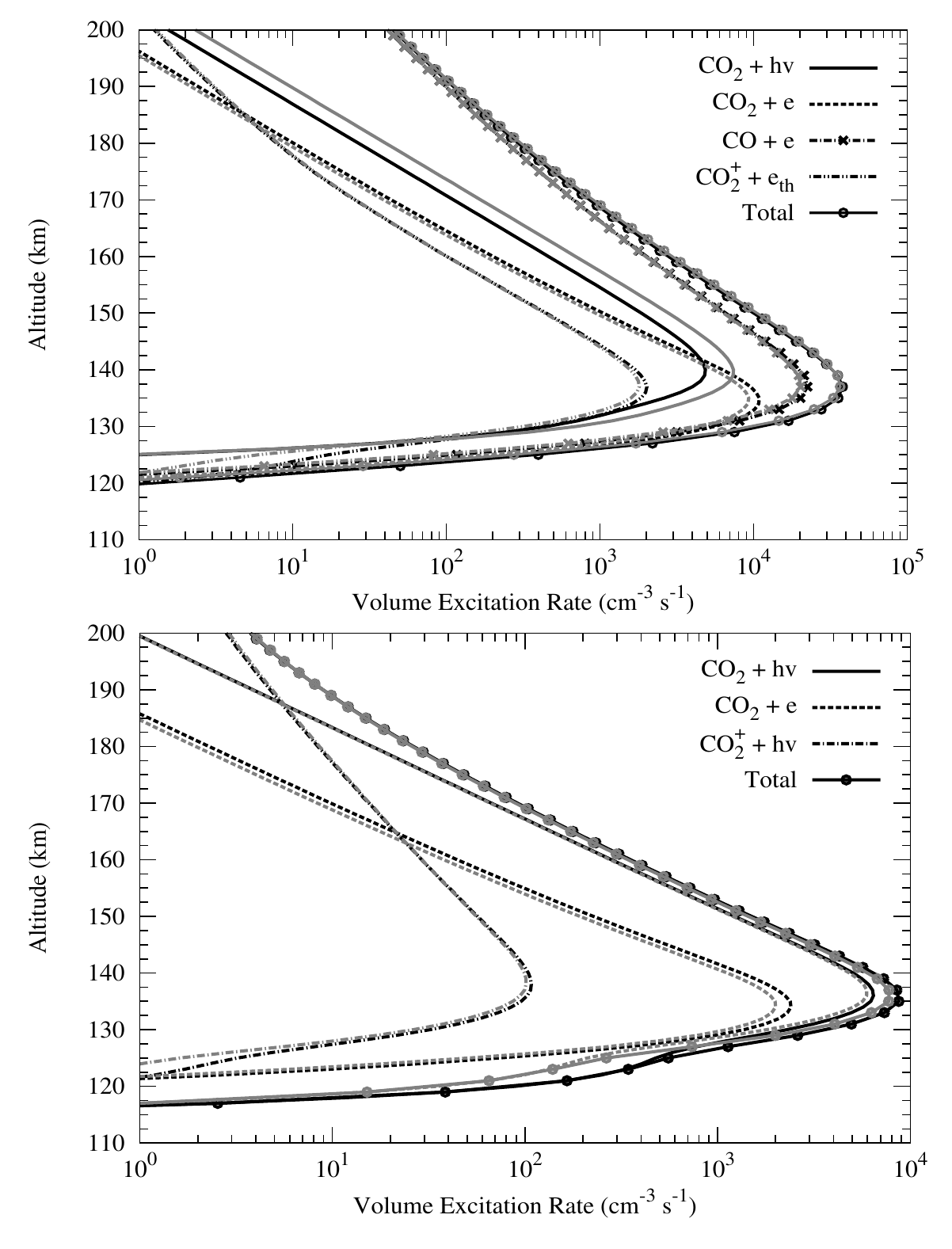}
\caption{Calculated production rates of the CO(a$^3\Pi$) (upper panel)  and CO$_2^+(\mathrm{B}^2\Sigma_u^+)$ (bottom panel)
for high solar activity condition  at SZA = 45$^\circ$. Black curves show calculated production rates using EUVAC
model while grey curve show them for S2K solar flux model.}
\label{fig:ver-max-venus}
\end{figure}

For solar maximum condition, \cite{Fox91} have
reported total Cameron band intensity of 57 kR, which is in 
agreement with the value of 60 kR in the present study. However, the
contribution of individual processes is different in the two studies. In the
present study the e-CO is the dominant process; whereas in the model calculation of \cite{Fox91}
the photon and electron impact on CO$_2$ played the dominant role with contribution
of about 36\% from each, while the contributions of electron impact on CO and
DR of CO$_2^+$ were 20 and 8\%, respectively.

The present study shows that the contribution of e-CO process in CO(a$^3\Pi$) production
is directly related to the cross section used in the calculation. For CO(a$^3\Pi$) cross section
of  \cite{LeClair94b}, the e-CO process is found  to be the dominant source
of CO Cameron band (see Figure~\ref{fig:ver-cameron-xs} and Table~\ref{tab:oi-cmp-venus}).
Overall, the calculation shows that the role of electron impact on CO in the Cameron band
production may have been underestimated in the earlier calculations of  \cite{Fox81} and
\cite{Gronoff08} due to the choice of e-CO cross section for CO(a$^3\Pi$) production used in
their calculations.

\begin{figure}
\centering
\includegraphics[width=20pc]{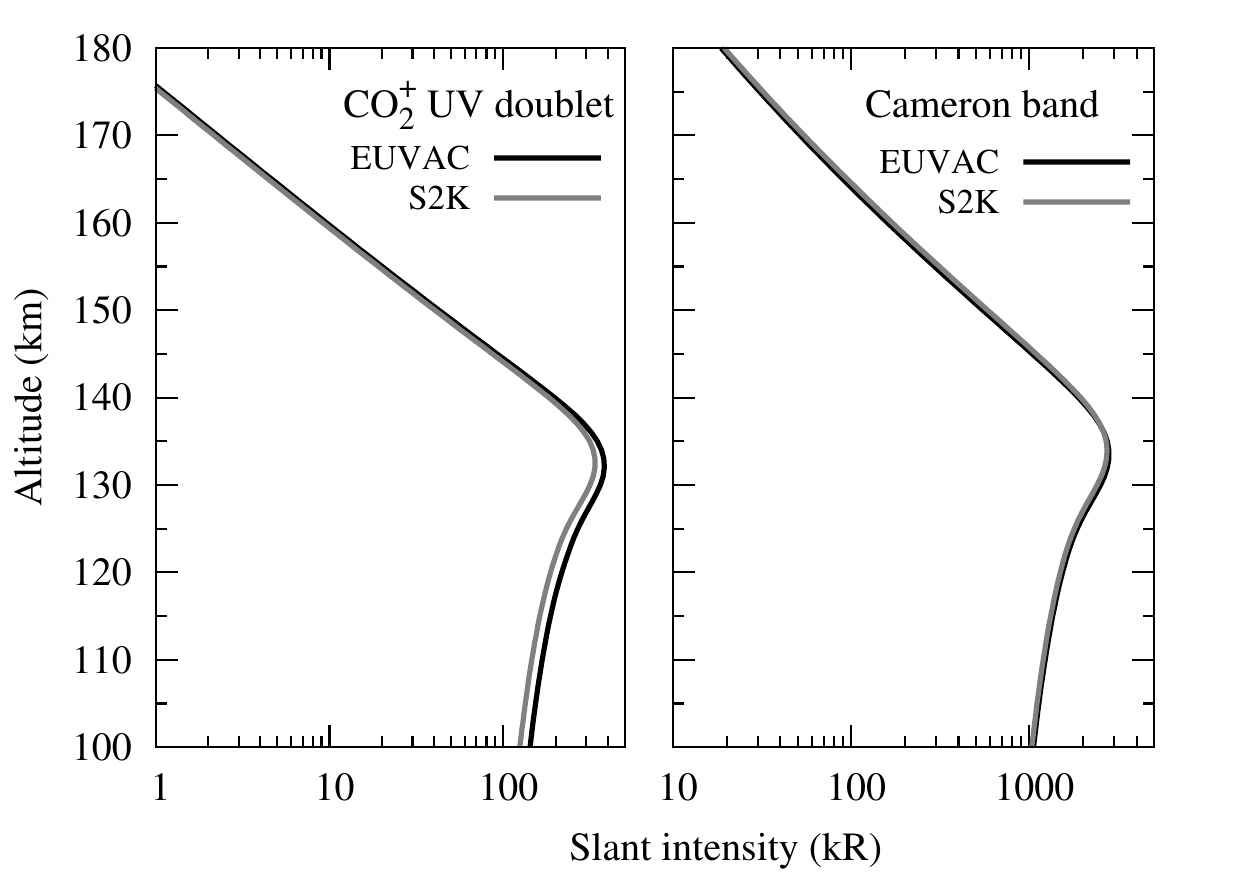}
\caption{Calculated limb profiles of CO$_2^+$ UV doublet (left panel) and
CO Cameron band (right panel)  for EUVAC and S2K solar EUV
flux models for high solar activity condition at SZA = 45$^\circ$.}
\label{fig:int-max-venus}
\end{figure}

It has been mentioned earlier that a branching ratio of 0.5 has been used to calculate
the UV doublet emission intensity because we have used excitation cross section of CO$_2^+$(B) in
the calculation.
However, if emission cross section of CO$_2$(B) given by \cite{Ukai92} is
used in the calculation rather than the excitation cross section,  the contribution of photoionization
in CO$_2^+$(B$ ^2\Sigma_u^+ $) ion production reduces by about 30\%. For example during 
solar minimum (maximum) condition the overhead intensity of UV doublet emission due to photoionization
of CO$_2$ is about 1.8 kR (2.8 kR). This value is about 25\% (40\%) smaller than that calculated
using CO$_2$(B) excitation cross section (if 50\% branching cross-over from B to A state is
considered  for excitation cross section). It shows that use of emission and excitation cross
sections of CO$_2$(B) affects the emission intensity of CO$_2^+$ UV doublet.

\subsection{Effect of solar EUV flux models}
During the solar minimum condition, the CO Cameron band excitation rate calculated
using the S2K model is about 45\% larger than that calculated using the EUVAC model, while
the production in  PD of CO$_2$ is about 50\% higher when
S2K model is used. However, the altitude of peak production is same 
for both solar EUV flux models (see Figure~\ref{fig:ver-low-venus}).
The limb intensities calculated using the S2K model are about 40\% larger than
those calculated using the EUVAC model (see Figure~\ref{fig:int-low-venus}).

\begin{figure}
\centering
\includegraphics[width=20pc]{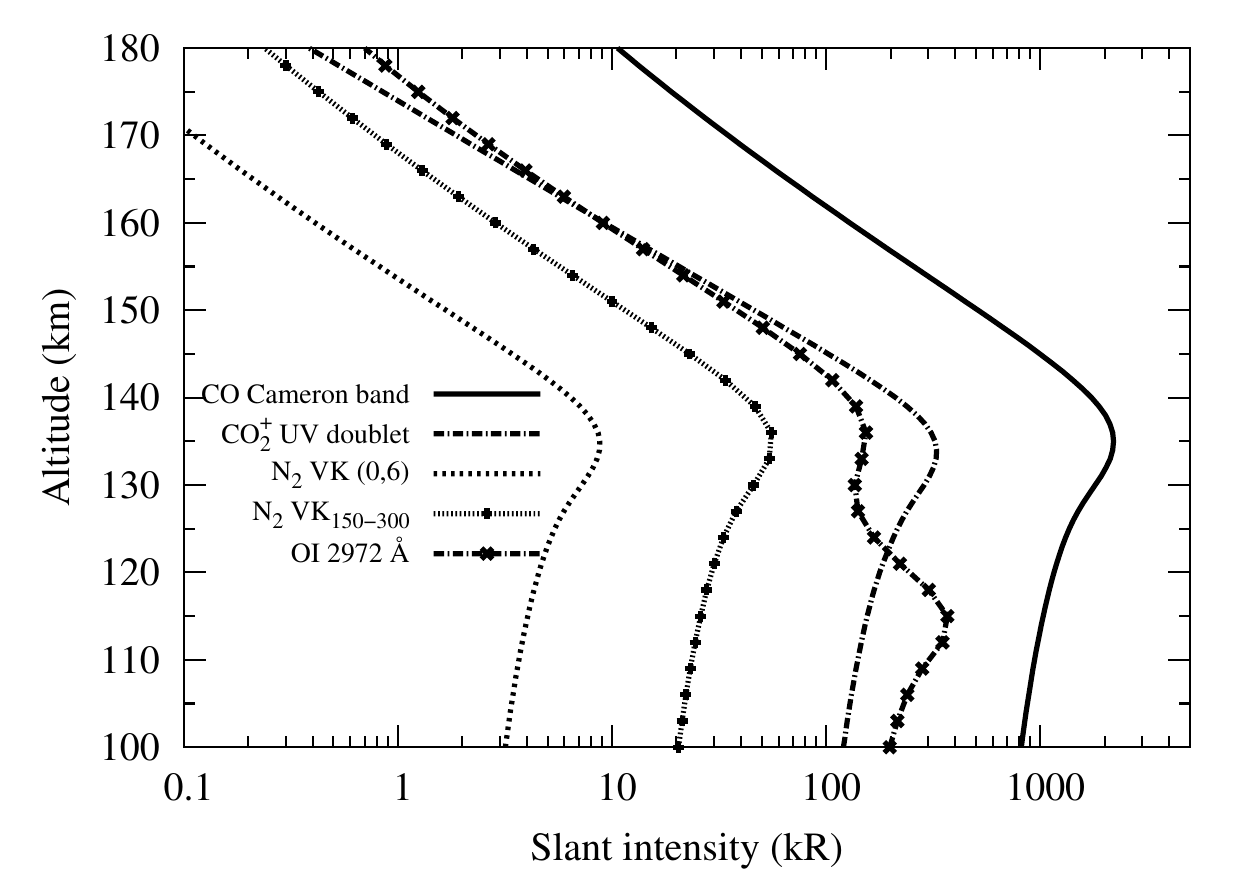}
\caption{The calculated (using EUVAC solar flux model) limb profiles
of CO$_2^+$ UV doublet  and CO Cameron band emissions for moderate solar activity condition  at SZA = 45$^\circ$.
The calculated limb intensities of OI 2972 \AA\ and N$_2$ VK (0, 6) emission are
also shown in the figure, along with the limb intensity of N$_2$ VK band in wavelength 
region 1500--3000 \AA.}
\label{fig:int-mod-venus}
\end{figure}

For high solar activity condition the intensity of CO Cameron band and CO$_2^+$ UV
doublet emissions calculated using the EUVAC model is about 2\% and 10\%,
respectively, higher than those calculated using the S2K model. This is due
to the higher solar EUV flux in EUVAC model at wavelengths $\le $250 \AA\  that
produces energetic photoelectrons which further ionize the medium and compensate
for the higher photoionization by solar EUV flux at wavelengths $>$250 \AA\ in the S2K model.
The effect of solar EUV flux on model calculations for moderate
solar activity condition is similar to that for high solar activity condition.
Similar variation in the emission intensities due to the change in EUV flux models
for solar minimum and maximum conditions have been found on Mars \citep{Jain12}.

\begin{figure}
\centering
\includegraphics[width=20pc]{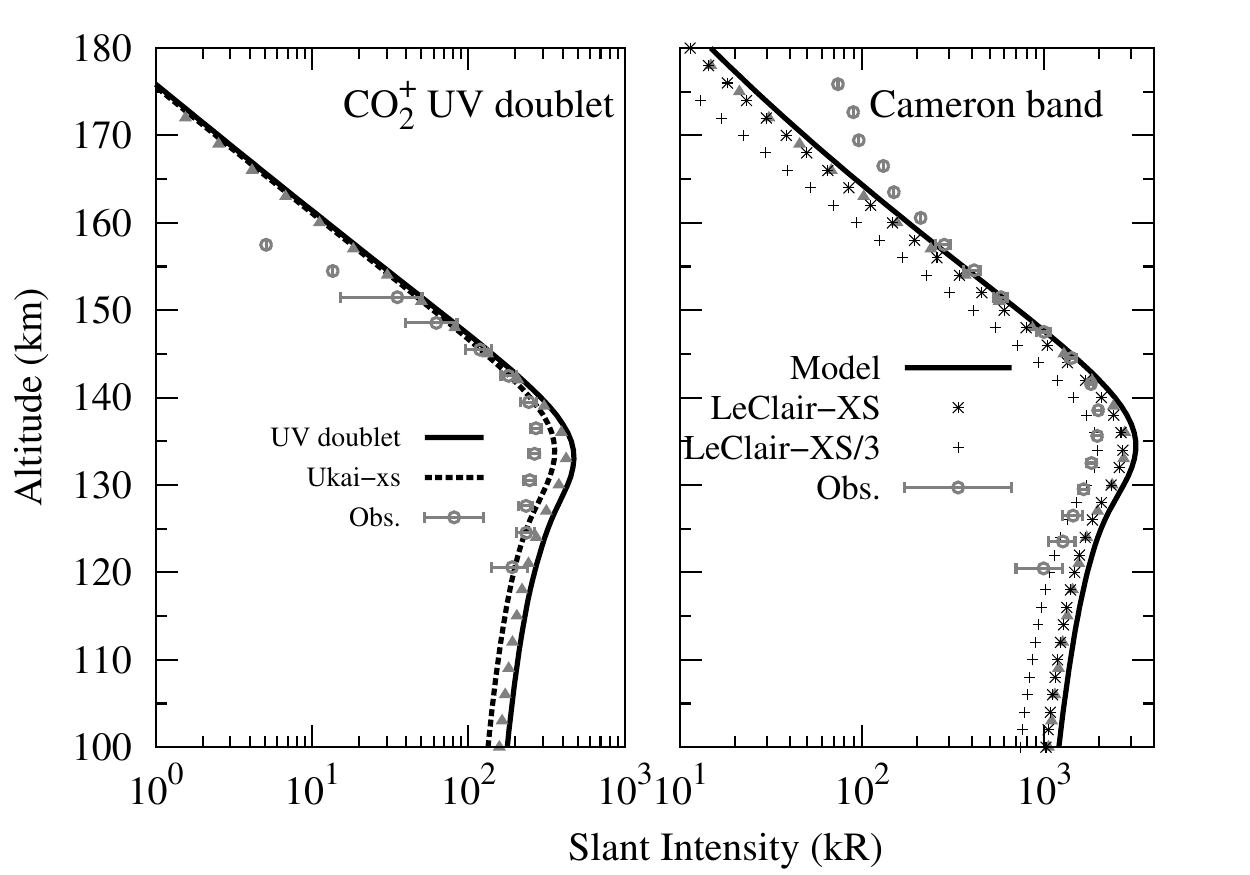}
\caption{The calculated (using EUVAC solar flux model) limb profiles
of CO$_2^+$ UV doublet  and CO Cameron band emissions for conditions similar to SPICAV observations
(F10.7 = 144 and SZA = 25$^\circ$) along with observed profiles taken from \cite{Chaufray12}. Limb
intensities of CO Cameron band, calculated by using the actual and corrected 
cross sections  of a$^3\Pi$ state in e-CO process measured by \cite{LeClair94b} are also shown. 
Solid triangles show the calculated intensity on 8 October 2011 (F10.7 = 118.3 and F10.7-81 days average = 
140.6, and SZA = 25$^\circ$). Limb intensity of CO$_2^+$ UV doublet calculated  by using  emission 
cross section of \cite{Ukai92}  is also shown (dashed curve).}
\label{fig:int-venus-spicav}
\end{figure}

For all the three (low, moderate, and high) solar activity conditions the contribution of
PD of CO$_2$ to the Cameron band production is 50\% higher when the S2K 
solar flux model is used. This is because of an order of magnitude higher solar EUV flux in
the 1000--1050 \AA\ bin in the S2K model compared to that in the EUVAC model. Solar EUV flux 
in the 1000-1050 \AA\ bin does not significantly contribute to the photoionization, but
mostly affects the PD of CO$_2$: thus affecting the Cameron band production
in the PD of CO$_2$.

For solar maximum condition, the calculated intensities using the EUVAC model
are two times higher than those calculated for solar minimum condition. When the S2K model
is used, the respective intensities of UV doublet and Cameron band emissions are 1.3 and
1.6 times larger in high solar activity than those in low solar activity
condition. For the EUVAC solar flux model, the variation in contribution of electron impact
processes are more prominent for change in solar activity from low to high due to a
change of more than a factor of 2 in the solar EUV flux below 250 \AA, whereas
solar EUV flux in the S2K model varies by less than a factor of 2 from 
solar minimum to maximum condition.

\subsection{CO($ a',d,e $) triplet emissions on Venus}\label{sec:co-visible}
The PD of CO$_2$ below 1080 \AA{} leads to the formation of CO(a$^3\Pi$), but at photon
energies greater than 12.4 eV (wavelength $<$ 1000 \AA), other channels open up.
The PD of CO$_2$ in the 10.3-13.8 eV (1200--900 \AA) region
leads to the channel CO* + O($^3$P), where CO* corresponds to four triplet levels
a$^3\Pi$, a$'^3\Sigma^+$, d$^3\Delta$, and e$^3\Sigma^-$. Emissions arising due to the 
transition from the $a'$,  $d$, and $e$ states to the a$^3\Pi$ state are called
Asundi, Triplet, and Herman bands, respectively. \cite{Conway81} reported that
the CO Cameron band spectra observed by Mariner showed a very hot rotational
distribution. His analysis showed a bimodal fit with temperatures 1600 K and
10,000 K. Analysis of SPICAM/Mars Express data also showed similar hot
distribution \citep{Kalogerakis12}.

Recently, \cite{Kalogerakis12} studied the PD of CO$_2$ in laboratory and found
strong emissions in the visible and near-IR region arising from the CO($ a',d,e $) triplet states.
They attributed these triplet band emissions to be the primary source
for the CO(a--X) Cameron bands. \cite{Kalogerakis12} concluded
that most of the observed Cameron band arising from PD of CO$_2$ is
preceded by the cascading from the CO($ a',d,e $) triplet states, and
predicted that the visible and near-IR (6000 to $ > $14000 \AA) emissions from
these triplet states is of the same magnitude as the CO Cameron band.

Using the study of \cite{Kalogerakis12}, one can predict the lower limit 
of Asundi, triplet, and Herman bands in the atmosphere of Venus,
if only PD of CO$_2$ is considered as the primary source of these
CO($ a',d,e $) triplet states. Results from the present study
show that for solar minimum condition the contribution of PD 
of CO$_2$ to the CO Cameron band production on Venus is 5.7 kR
(see Table~\ref{tab:oi-cmp-venus}). Thus, the CO($ a',d,e $) emissions would also be
about 5.7 kR on Venus, spread over the 6000 to $>$14000
\AA{} range. The Asundi $ a'-a $ (5-0) band at 7830 \AA{} is about 10\% of the total
triplet band emissions \citep{Kalogerakis12}, thus its overhead intensity
on  Venus would be about 570 R. Similarly, during the
solar maximum condition total intensity of CO($ a',d,e $) triplet and
Asundi $ a'-a $ (5-0) bands on Venus would be 7.5 kR
and 750 R, respectively. The maximum fraction of Cameron band originates from electron
impact on CO$_2$ and CO on Venus and these processes do not exclude similar CO product
\citep{Kalogerakis12}. The magnitude of CO($ a',d,e $) triplet
bands on Venus reported above would be a lower limit; hence,  an upper
limit could be larger by a factor of 2 to 3.

\subsection{Calculation of other ultraviolet emissions}
Currently, after a prolonged minimum, the Sun is in the ascending phase of solar activity with
moderate condition. As discussed in Section~\ref{subsec:smod-venus}, we have carried
out model calculations for the moderate (F10.7 = 130) solar activity condition.
For the SPICAV/VEX observation of UV dayglow emissions during the current solar moderate condition,
our model predicts the CO Cameron band (CO$_2^+$ UV doublet) intensity of $\sim$2200 kR (330 kR)
at an altitude of $ \sim $135 km. Based on our earlier
calculations of N$_2$ triplet band emissions on Venus \citep{Bhardwaj12b}, 
in moderate solar activity condition we predict the maximum 
intensity of about 10 kR for N$_2$  Vegard-Kaplan (0, 6) emission at the altitude of
135 km (see Figure~\ref{fig:int-mod-venus}).
The N$_2$ VK (0, 6) emission at 2762 \AA\ has been observed on Mars \citep{Leblanc06}, and
is the brightest emission in the N$_2$ VK band system \citep{Jain11,Bhardwaj12b}. The intensity of
other prominent transition of N$_2$ VK band can be calculated using the
intensity ratio provided in our earlier calculations \citep{Jain11,Bhardwaj12b,Bhardwaj12c}.
We have also calculated the limb intensity of N$_2$ VK band in the wavelength range
1500--3000 \AA\ (that lies within the SPICAV UV measurement range), which is shown in 
Figure~\ref{fig:int-mod-venus}. The maximum limb intensity of 
N$_2$ VK band in the 1500--3000 \AA\ range  is about 60 kR for solar 
moderate activity condition.

We have recently developed a model for visible atomic oxygen dayglow emissions
in the atmosphere of Mars [Jain and Bhardwaj, in preparation]. We have applied this model on Venus and calculated
the atomic oxygen 2972 \AA\ (which is within the SPICAV UV measurement range) emission on Venus for
moderate solar activity  condition. The calculated limb profile of OI 2972 \AA\ emission
is presented in Figure~\ref{fig:int-mod-venus}, which shows two peaks: the lower peak at
$ \sim $115 km has intensity of 375 kR, while the upper peak at $\sim$135 km has intensity of
154 kR. The upper peak is mainly due to the photodissociation
of CO$_2$ at wavelengths between 860 and 1160 \AA, while the lower peak is due to PD of CO$_2$
by solar H Ly-$\alpha$ photons (1216 \AA). Recent analysis of SPICAM-observed OI 2972 \AA\ emission profile
on Mars also suggests a double peak structure \citep{Gronoff12}.

\section{Comparison of model calculations with the recent SPICAV observation}
Within  weeks of submitting this manuscript, \cite{Chaufray12} reported
the first dayglow observation of CO Cameron band and CO$_2^+$ UV doublet emissions
on Venus by SPICAV aboard Venus Express. The SPICAV observations
were made between October and December 2011, with  solar zenith angles varying between 20$^\circ$ and 
30$^\circ$. We have carried out calculation for the similar condition
as reported by \cite{Chaufray12} by taking SZA of 25$^\circ$ and VTS3 model atmosphere
for 15 November 2011 (F10.7 = 148 and F10.7-81 days average = 144).
Figure~\ref{fig:int-venus-spicav} shows the calculated CO Cameron band and CO$_2^+$ UV 
doublet brightness profiles along with the SPICAV-observed profiles taken from
\cite{Chaufray12}.

The model calculated brightness of CO Cameron band peaks at 134 km with a value of 3200 kR.
The SPICAV-observed peak of Cameron band brightness is situated 
at $137\pm1.5$ km and the magnitude of limb intensity at this altitude is
$\sim$2000 kR \citep{Chaufray12}. The calculated intensity at the peak is about 50\% higher
than the observed value. When the CO(a$^3\Pi$)  production cross section  in e-CO collision
of \cite{LeClair94b} is used, the limb intensity of Cameron band at the peak altitude
is 2700 kR. As mentioned earlier, the cross section obtained by
\cite{LeClair94b} might be overestimated by a factor of 3 (see Section~\ref{sec:model}).
On decreasing the \citeauthor{LeClair94b}'s measured cross section by a factor of 3, the
calculated CO Cameron band brightness at the peak is $\sim$2000 kR.
For CO$_2^+$ ultraviolet doublet emission, maximum limb intensity  of $\sim$470 kR is obtained
at an altitude of 133 km, which is $ \sim $70\% higher than the SPICAV-observed value of 270
kR (at $ 135.5\pm2.5 $ km) \citep{Chaufray12}. However, this difference is maximum at peak only.
At altitudes above (below) the peak, say at 150 km (120 km), the calculated intensity is in agreement
with the SPICAM observation. If emission cross section of CO$_2$(B$ ^2\Sigma_u^+ $) given by
\cite{Ukai92}  is used in
the calculation then calculated UV doublet emission intensity (see dashed curve in 
Figure~\ref{fig:int-venus-spicav}) is $\sim$30\% higher than the
observation. 
The observed profile of CO$_2^+$ UV doublet emission may contain a small portion
of OI 2972 \AA\ emission \citep{Chaufray12},
which makes the shape of observed brightness profile different than the calculated emission
profile at lower altitudes.
The comparison between calculation and observation depends
on factors such as  local variations in the neutral atmosphere--density and temperature depending on F10.7, 
winds and vertical transport, averaging over 3 months to get the adequate S/N ratio for the observational
profiles, and moreover uncertainty in the model calculation.
For example, model calculated intensities of CO Cameron band and   CO$_2^+$ UV doublet emissions
decreases by $ \sim $13\%  (see Figure~\ref{fig:int-venus-spicav})
on 8 October 2011 (F10.7 = 118.3 and F10.7-81 days average = 140.6, and SZA = 25$^\circ$).

\cite{Chaufray12} have derived the overhead intensity of 25.3 kR and 3.2 kR for Cameron band
and CO$_2^+$ UV doublet emissions, respectively, by converting the limb intensity to zenith
brightness above sub-solar point. These values are significantly lower than our model calculated
height-integrated overhead intensities of  70 and 8 kR (at SZA = 25$^\circ$) for Cameron band
and CO$_2^+$ UV doublet emissions, respectively.
This discrepancy in the calculated and observation-derived overhead intensity is significant
and it is difficult to reconcile or comment on the cause for this difference at present and further
investigation is needed.

The calculated  altitude of peak brightness of both CO Cameron band and CO$_2^+$ UV doublet
emissions is  lower by $ \sim $5 km than the observation. The difference in peak altitude 
of observed and calculated emissions shows that the upper atmospheric neutral density is smaller
in our model calculation. Recent  general circulation model for Venus (VTGCM) also
suggests that VTS3 empirical model is inadequate to properly represent lower
thermosphere thermal structure \citep{Brecht12}. Density profile of CO$_2$ calculated by VTGCM
vary significantly from that calculated by  VTS3  model above 100 km.

\section{Summary and Conclusions}\label{sec:summary}
We have presented the model calculation of CO Cameron band and CO$_2^+$
doublet ultraviolet emissions in the dayglow of  Venus and assessed the impact of
solar EUV flux model on the calculated intensities. The calculated
volume production rates of CO Cameron band and CO$_2^+$ UV doublet emissions  are height-integrated
to compute the  overhead intensity and integrated along the line of sight to obtain the limb
intensities for low, moderate, and high solar activity conditions.
With updated cross section, the electron impact on CO is found to be 
the major source of CO(a$^3\Pi$) 
production followed by electron and photon impact dissociation of CO$_2$.
The major source of CO$_2^+$ UV doublet emission in Venusian dayglow
is photoionization of CO$_2$ followed by electron impact ionization of CO$_2$.
The contribution of fluorescence scattering by CO$_2^+$ to the  CO$_2^+$ UV doublet
emission is quite negligible. The calculated overhead intensities of CO Cameron band and CO$_2^+$ UV
doublet emission are about a factor of 2 higher in the solar maximum
condition than those during the solar minimum condition. This variation
in intensity from low to high solar activity depends upon the solar
EUV flux model used in the calculation, e.g., when the S2K model is used instead
of EUVAC, the emission
intensities of CO Cameron  band and CO$_2^+$ UV doublet vary by less than 
a factor of 2. The effect of solar EUV flux models on the emission intensity 
is 30-40\% in solar minimum condition  and $ \sim $2-10\% in solar
maximum condition.

For the SPICAV/VEX observation of UV dayglow emissions during  
the solar moderate condition, we have predicted the limb intensity
of about 2400 and 300 kR  for CO Cameron band  and CO$_2^+$ UV doublet emissions,
respectively. We have also predicted the intensities of N$_2$ Vegard-Kaplan UV bands
($ \sim $60 kR in wavelength range 1500--3000 \AA, peaking at $\sim$ 135 km) and OI 2972 \AA\ emission
(375 kR at lower ($ \sim $ 115 km) and 155 kR at upper ($ \sim $135 km) peak) in moderate solar activity condition.

We have compared our calculated limb intensity of CO Cameron band 
and CO$_2^+$ UV doublet emissions with the first observation of these emissions
on Venus using SPICAV/VEX \citep{Chaufray12}. The calculated intensity of CO Cameron band
at the peak altitude is about 50\% higher
than the SPICAV observation. However, when the CO(a$^3\Pi$) production  cross section in e-CO collision
measured by \cite{LeClair94b} is used in the model calculation, this difference reduces
to 30\% and with a correction by a factor of 3 in cross section, the magnitude of
calculated brightness at peak is in good agreement with the observation.
The calculated maximum brightness of CO$_2^+$ doublet emission is $\sim$70\% higher
than the SPICAV observation. However, when CO$_2^+$(B) emission cross section of \cite{Ukai92}
is used, the calculated maximum intensity agrees better with the observation.
We  found that our calculated overhead intensities
of the two emissions is significantly higher than those derived from the 
observations. It may be noted that a number of factors can affect the comparison between
observation and calculation, e.g., observed brightness profiles are averaged of several measurements spanning
over 3 months during which variation in the solar zenith angle and other local variations
in neutral atmosphere and temperature  can affect the dayglow emissions. Moreover, uncertainties
in the model input parameters can also cause  discrepancy between observed and calculated
brightness profiles.
Presently, it is difficult to comment on this discrepancy and further investigation is needed.
Our model calculated peak altitude of CO Cameron
band and CO$_2^+$ UV doublet emission profiles is lower than that observed by
SPICAV, indicating lower neutral density in the VTS3 model atmosphere for Venus used in our calculation.

The present study has clearly demonstrated that the cross section of a$^3\Pi$ state in e-CO process
is important in modelling CO Cameron band emission on Mars and Venus. 
The contribution of e-CO process in CO Cameron band  also depends on the 
density of CO in the atmosphere; hence, it is difficult to constrain the former
without fixing the latter. Present calculation also showed that use of excitation and
emission cross section of CO$_2^+$(B) can affect the UV doublet emission intensity, 
and one should be careful while using these cross sections in the model calculation.
A more detailed study of these emissions taking the Venus thermosphere 
general circulation model (VTGCM) needs to be carried out to understand the recent SPICAV observations.

\newpage

\renewcommand{\thefootnote}{\fnsymbol{footnote}}

\clearpage

\clearpage
\newpage

\end{document}